\documentclass{svproc}
\usepackage{url}

\usepackage{graphicx}
\usepackage{amsmath}
\usepackage{amssymb}
\begin{document}
\mainmatter              
\title{Maxwell's Demon: Controlling Entropy via Discrete Ricci Flow Over Networks}

\author{Romeil Sandhu\inst{1}, Ji Liu \inst{1}}

\institute{Stony Brook University, Stony Brook NY 08544, USA,\\
\email{romeil.sandhu@stonybrook.edu}}

\maketitle              

\begin{abstract}
In this work, we propose to utilize discrete graph Ricci flow to alter network entropy through feedback control.   Given such feedback input can ``reverse'' entropic changes, we adapt the moniker of Maxwell's Demon to motivate our approach.  In particular, it has been recently shown that Ricci curvature from geometry is intrinsically connected to Boltzmann entropy as well as functional robustness of networks or the ability to maintain functionality in the presence of random fluctuations.  From this, the discrete Ricci flow provides a natural avenue to ``rewire'' a particular network's underlying geometry to improve throughout and resilience.  Due to the real-world setting for which one may be interested in imposing nonlinear constraints amongst particular agents to understand the network dynamic evolution, controlling discrete Ricci flow may be necessary (e.g., we may seek to understand the entropic dynamics and curvature ``flow'' between two networks as opposed to solely curvature shrinkage).  In turn, this can be formulated as a natural control problem for which we employ feedback control towards discrete Ricci-based flow and show that under certain discretization, namely Ollivier-Ricci curvature, one can show stability via Lyapunov analysis.  We conclude with preliminary results with remarks on potential applications that will be a subject of future work.
\end{abstract}
\section{Introduction}
In the current technological world, we increasingly depend upon the reliability, robustness, quality of service and timeliness of exceedingly large interconnected dynamical systems including those of power distribution, biological, transportation, and communication \cite{Bara1}.  Over the past twenty years, we have witness a dramatic rise of information in which the analysis of such systems invariably present challenging ``big data'' complexity issues.  For example, in transferring resources and information, a key requirement is the ability to adapt and reconfigure in response to structural and dynamic changes while avoiding disruption of service. In turn, exploiting functional properties such as robustness and heterogeneity (redundancy) are key to maintaining control and avoiding shotgun-based solutions during ``black swan'' events in which the continuous failing of interacting agents may result in catastrophic system failure.  
\begin{figure}[!t]
\begin{center}
\includegraphics[width=1\textwidth]{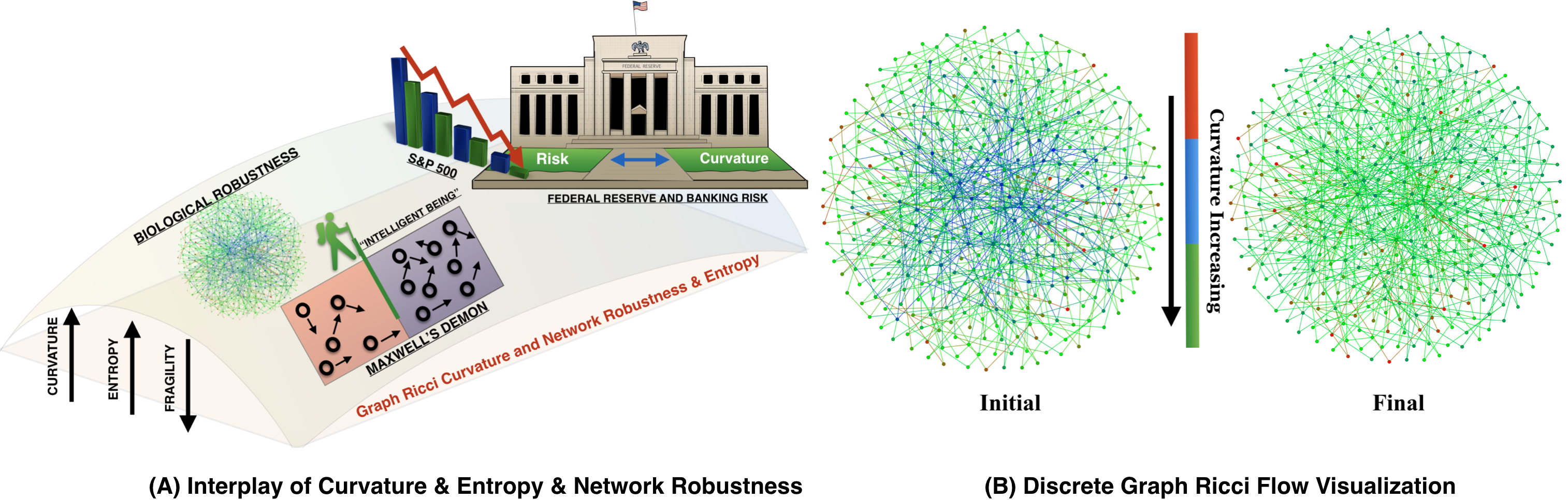}
\caption{Motivated by Maxwells ``Demon'', this work focuses on altering network entropy via Ollivier-Ricci flow whereby the ``intelligent being'' is a feedback operator.}
\label{fig:sweet_spot}
\end{center}
\end{figure}

As such, we have previously developed fundamental relationships between network functionality \cite{Dem1,Varadhan} and certain topological and geometric properties of the corresponding graph \cite{Ollivier1,Ollivier2} to show that the geometric notion of curvature (a measure of ``flatness'') is positively correlated with network entropy and system's robustness or its ability to adapt to dynamic changes \cite{Sandhu1,Sandhu2}.  This can be seen in Figure \ref{fig:sweet_spot}. In this regard, network curvature may relate to anomaly detection, congestion in communication, to drug resistance.   On the other hand, network entropy has often been chosen as a measure of network functional robustness \cite{Dem1,Tesch1}. From this, if one is able to define such statistical properties over the graph that are proxies for functionality, then a natural progression would be to define corresponding theoretics in order to alter the networks behavior through such properties and for which in this note, we consider curvature and entropy. To this end, we focus on developing the necessary conditions to control network (curvature) entropy through the discrete Ricci flow.  This flow in the graph setting has been proposed for congestion management, managing systemic risk \cite{Sandhu2}, simulating biological resistance \cite{Sandhu1}, as well as a generalized tool for network comparison \cite{Emile1,Emile2}.  This said, the discrete Ricci flow for networks presents notable issues in that it not only reduces regions of negative curvature, but also reduces areas of highly positive curved regions.  In the context of inducing network fragility (or vice versa), this may not be suitable as \emph{increases} in negatively curved regions curvature relates to increases in entropy and subsequently network robustness. Further, in understanding network dynamics, one may want to ``drive'' the discrete flow between two networks \cite{Emile1,Emile2} as well as in augmented fashion for which one ``pins down'' the flow on regions considered ``undruggable.'' The remainder of this note is outlined as follows: The next section provides preliminaries in motivating the theoretical need of understanding geometry as it pertains to functionality. From this, Section 3 lays the foundation of our framework for which we present the corresponding control laws and prove stability in the sense of Lypanuv.  Then, Section 4 presents  preliminary results on synthetic networks for illustration of theory.  We conclude with a summary and future work towards applications in Section 5.

\section{Preliminaries:  Entropy and Curvature}
To illustrate how geometry elucidates the functional behavior of a dynamical system, let us revisit optimal mass transport (OMT) \cite{Villani}. The first notion of OMT was proposed by Gaspar Monge in 1781 with the concern of finding the minimal transportation cost for moving a pile of soil from one site to another. The modern formulation, given by Kantorovich, has been ubiquitously used in fields of econometrics, fluid dynamics, to shape analysis \cite{Villani,Brenier} and recently, has received a renewed mathematical interest. More formally, let $(X,\mu_0)$ and $(Y,\mu_1)$ be two probability spaces and let $\pi(\mu_0,\mu_1)$ denote the set of all couplings on $X\times Y$ whose marginals are $\mu_0$ and $\mu_1$.  As such, the Kantorovich costs seeks to minimize $\int c(x,y) d\pi(x,y) \forall \pi\in \pi(\mu_0,\mu_1)$ where $c(x,y)$ is the cost for transporting one unit of mass from $x$ to $y$.  The cost originally defined in a distance form on a metric space leads to the $L^p$ Wasserstein distance as follows:
\begin{equation}
W_p(\mu_0,\mu_1):= \bigg(\inf_{\mu\in\pi(\mu_0,\mu_1)} \int\int d(x,y)^pd\mu(x,y))\bigg)^{\frac{1}{p}}.
\end{equation}
From this, let us begin considering $M$ to be a Riemannian manifold such that
\begin{align}
\label{eq:scalar_structrure}
\begin{split}
&\mathcal{P} := \big\{ \mu\geq 0: \int \mu \text{ dvol}(M)=1 \big \} \quad\quad \mathcal{T_\mu P} := \big\{ \eta : \int \eta \text{ dvol}(M)=0 \big \}
\end{split}
\end{align}
as the space of probability densities and the tangent space at a given point $\mu$, respectively.  Due to the work of Benamou and Brenier \cite{Brenier}, one can naturally compute the geodesic (\emph{in the Wasserstein sense}) between two densities $\mu_0,\mu_1\in \mathcal{P}$ 
as the below optimal control problem:
\begin{align}
\begin{split}
\inf_{\mu,g} &\bigg\{\int\int_{0}^{1} \mu(t,x) \rVert \nabla g(t,x) \lVert dt \text{dvol}(M) \\ 
&\text{subject to} \quad \frac{\partial u}{\partial t} + \text{div} (\mu\nabla g) = 0  \\
&\quad\quad \mu(0,.)=\mu_0, \quad \mu(1,.)=\mu_1 \bigg\} 
\end{split}
\label{eq:brenier}
\end{align}
which leads us to give $\mathcal{P}$ a Riemannian structure due to the work of Jordan \emph{et. al} \cite{Jordan}.  From this, we can now consider Boltzmann entropy as
\begin{equation}
H(\mu_t):= \int_M \log \mu_t \text{dvol}(M)
\end{equation}
where the dependency on $x$ has been dropped for convenience and we consider a family of densities evolving over time.  Taking the second variation with respect to time $t$ in the Wasserstein sense (i.e., rather than the Euclidean norm) and noting that, by construction, $\eta:=\frac{\partial \mu}{\partial t}|_t=0$, we have  
\begin{align}
&\frac{d^2}{dt^2}H(\mu_t)|_{t=0} = \langle Hess(H)(\eta),\eta\rangle_W \nonumber \\
&= -\int_M \langle \nabla g_\eta,\nabla \Delta g_\eta\rangle +\frac{1}{2}\Delta\big(\lVert \nabla g_\eta\rVert^{2}\big)\mu_0 \text{dvol}(M)
\end{align}
where $\mu_0$ and $g_\eta$ satisfy (\ref{eq:brenier}).  Using the Bochner formula \cite{Chow}, which relates harmonic functions on a Riemannian manifold to Ricci curvature (herein denoted as ``Ric''), we can further assume $Ric \geq kI$ as quadratic forms where $k$ is a constant and $I$ is the identity matrix.  Then, due to Sturm \cite{Sturm} as well as Lott and Villani \cite{Villani}, one can show that the $Hess(H)$ is k-convex:
\begin{equation}
H(\mu_t)\!\!\leq t H(\mu_0)\! +\!(1-t)H(\mu_1) - \phi(k,t,\mu_0,\mu_1) \forall t\in[0,1] 
\label{eq:entropy_curv_result}
\end{equation}
where the right hand portion $\phi(.)$ can be shown to be $\phi(k,t,\mu_0,\mu_1)= \frac{k}{2} t(1-t)W_2 (\mu_0,\mu_1)^2$ allowing for k-convexity.  That is, changes in entropy and curvature are positively correlated, i.e., $\Delta H \times \Delta Ric \geq 0 $.   Furthermore, through the Fluctuation Theorem \cite{Dem1}, one may relate network robustness $R$ to entropy; i.e., $\Delta H \times \Delta R \geq 0$ as well as Ricci curvature $\Delta Ric \times \Delta R \geq 0$ - see \cite{Dem1,Sandhu1,Sandhu2} for details.

\section{Proposed Framework}
In this section, we propose a feedback based approach to control discrete Ricci flow over graphs due to a discretization by Ollivier \cite{Ollivier1,Ollivier2} which is discussed next.
\subsection{Open-Loop View:  Discrete Ollivier-Ricci Flow}
While Ricci curvature relates to functionality, we require a discrete definition for networks.  Here, we focus on the Ollivier formulation \cite{Ollivier1} given its relationship to the Wasserstein distance, but refer to the reader to several works in this open problem area of varying discretizations including, but not limited to, Forman curvature \cite{Emile2,Emile3}, Bakery Emery \cite{AllenCompare} as well as recent comparisons \cite{Emile5,AllenCompare}. This said, we can define Ollivier-Ricci curvature between any two nodes $x$ and $y$ as:
\begin{equation}
\kappa(x,y) := 1 - W_1(\mu_x,\mu_y)/d(x,y).
\label{eq:Ollivier}
\end{equation}
This definition, motivated by coarse geometry, is applicable to the graph setting whereby the geodesic distance $d(x,y)$ is given by the hop metric. From this, we can define the Ollivier-Ricci flow with an initial condition $\mu_0(x,y) = \phi_0(x,y)$ as:
\begin{align}
\frac{d}{dt}\mu_t(x,y) :&= -\kappa(x,y)\mu_t(x,y) 
\label{eq:OllivierFlowOpen}
\end{align}
where $\mu_t(x,y)$ (with an abuse of notion) is the normalized edge weights, i.e., $\mu_t(x,y)\in [0,1]$. Here, we can treat this flow as an open-loop control problem \cite{AllenBook} for which the ``dynamics'' to be controlled is Ollivier-Ricci curvature $\kappa(x,y)$.  In particular, motivated philosophically by Maxwell's Demon \cite{Demon}, we seek to characterize an ``intelligent being'' to control entropy via discrete Ricci flow. 
\subsection{Control Law Construction and Existence}
To begin developing our control-based approach, let us redefine the above flow as a closed-loop problem with the following form given as:
\begin{align}
\frac{d}{dt}\mu_t(x,y) &= \big[ -\kappa(x,y)+\psi(\mu_t,\mu^*)\big]\mu_t(x,y) \\
\mu_0(x,y) &= \phi_0(x,y) \nonumber
\label{eq:OllivierFlowClosed}
\end{align}
where $\lim_{t\rightarrow \infty} \mu_t(x,y)\rightarrow \mu^*(x,y)$ and where $\psi(\mu,\mu^*)$ is the control law whereby the system is stable in the sense of Lypanuv (e.g., inputs ``near'' equilibrium stay or decay towards equilibrium).   Here, we assume $\mu^*(x,y)$ is ideal and for which there exists no error; i.e., we want the flow to converge entirely to such an end (network) point.   To do so, let us further define the point-wise and total error as
\begin{align}
\delta_t(x,y) & := \mu_t(x,y) - \mu^{*}(x,y)  \\
\Sigma_t (\delta_t) &:= \frac{1}{2} \sum_x\sum_y ||\delta_t(x,y)||^2.
\label{eq:ErrorAutonomous}
\end{align}
Given this, we are now able to show the existence of a regulatory control.
\\

\noindent\textbf{Theorem III.1}. \emph{The control law that stabilizes the closed-loop system in equation (9) from $\mu_t(x,y)$ to  $\mu^*(x,y)$ is given by:}
\begin{align}
\psi(\mu,\mu^*) & = \beta_t^2 (x,y) \delta_t(x,y)
\label{eq:ControlLaw}
\end{align}
\emph{where $\beta^2(x,y)\geq 2$} and $\delta_t(x,y)$ is given by equation (\ref{eq:ErrorAutonomous}).
\\

Proof: Let us first note that $\delta_t(x,y)$ is bounded, i.e., $-1\leq \delta_t(x,y) \leq 1$ and that $-2\leq \kappa(x,y) \leq 1$.  From this, we choose $\Sigma_t$ as the candidate Lyapunov function and differentiate it with respect $t$ which yields the following:
\begin{align}
\frac{d \Sigma_t}{dt} &= \sum_{x,y} \delta_t(x,y)\!\cdot\!\frac{\partial \delta_t(x,y)}{\partial t}           \nonumber \\
&=  \sum_{x,y}	\delta_t(x,y)\!\cdot \!\big [\frac{\partial}{\partial t}\mu_t(x,y) - 	\underbrace{\frac{\partial}{\partial t}\mu^*(x,y)}_0\big ]															      \nonumber \\		
&=  \sum_{x,y}	\delta_t(x,y)\!\cdot\!\big [-\kappa(x,y) - \beta_t^2(x,y)\delta_t(x,y)\big ] \mu_t(x,y)																 \nonumber \\
&\leq  \sum_{x,y}	\big[ |\kappa(x,y)|\delta_t^2(x,y) -\beta_t^2(x,y)\delta_t^2(x,y)\big ]	\mu_t(x,y)																 \nonumber \\
&\leq    \sum_{x,y}\delta_t^2(x,y)\big[ 2 -\beta_t^2(x,y)\big ]	\mu_t(x,y)																				 \nonumber \\
&\leq    0																		 \nonumber 
\label{eq:ProofI}
\end{align}
We note that while the above control law is due to the reliance on bounds of Ollivier-Ricci curvature, this will not hold for other discretizations such as Forman curvature \cite{Emile1,Emile2}. We have also assumed that one not only has an ideal representation of the corresponding network configuration $\mu^*(x,y)$, but the input is error-free and there are no modifications by a ``demonic'' operator during the entropic (Ricci) flow, e.g., impose node constraints.   This is discussed next.
\subsection{``Non-Perfect Demonic'' Input}
We are now ready to define an estimator and observer-like framework for which an input may begin to control graph curvature and subsequently control network entropy.   This can be akin to the thought experiment proposed by James Maxwell for which the ``demon'' seeks to violate the second law of thermodynamics, namely alter entropy \cite{Demon}.  Here, we assume there exists error from \textbf{both} the demon (and end targeted) state as well as the chosen (Ollivier-Ricci) flow model.  As such, let us define $\hat{\mu}_t^*(x,y)$ as the estimate of the ideal knowledge $\mu^*(x,y)$ with corresponding error terms associated with the demon and the model as
\begin{align}
\hat{\delta}_t(x,y) & := \mu_t(x,y) - \hat{\mu}_t^*(x,y) 		\quad\quad	\text{             (Type I Error)} \nonumber \\
\gamma_t(x,y) &:=	\hat{\mu}_t^*(x,y) - \lambda_t(x,y) 	\quad\quad						 \text{             (Type II Error)} \nonumber 
\end{align}
where $\lambda_t(x,y) := \sum_{l=0}^{l=k} \epsilon_t^k(x,y)$ and $\epsilon_t^k(x,y)  := \pm p \text{  (constant)}$ are the $k$ input at time $t$. From this, 
the total error for the above Type I / II errors can be seen as:
\begin{align}
\hat{\Sigma}_t(x,y)&:= \frac{1}{2}\sum_x\sum_y ||\hat{\delta}_t(x,y)||^2 \\
\Gamma_t(x,y) & := \frac{1}{2} \sum_x\sum_y |\lambda_t(x,y)| ||\gamma_t(x,y)||^2 .
\end{align}
\\
\noindent\textbf{Theorem III.2}.  \emph{Let us assume input has stopped and further assume the above total label errors defined for Type I/II error, then the following flow}
\begin{align}
\frac{d}{dt}\hat{\mu}_t(x,y) &= \big[ \hat{\delta}_t(x,y)+ \Phi(\lambda_t,\gamma_t)\big]\hat{\mu}_t(x,y) \\
\hat{\mu}_0(x,y) &= \phi_0(x,y) \nonumber
\end{align}
\emph{where $\Phi(\lambda_t,\gamma_t) = -|\lambda_t(x,y)|\gamma_t(x,y)$ provides an estimator such that the total error $V_t(x,y) :=  \hat{\Sigma}_t(x,y)+\Gamma_t(x,y)$ has a negative semi-definite derivative. In turn, this provides a stable coupled feedback system together with equation (9) where the ideal configuration $\mu^*(x,y)$ is replaced with an estimator $\hat{\mu}_t^*(x,y)$.}
\\

Proof:  Computing the total error $V_t(x,y) :=  \hat{\Sigma}_t(x,y)+\Gamma_t(x,y)$ and dropping the spatial dependency (for reading ease), yields the following:
\begin{align}
\frac{\partial V_t}{\partial t} &= \sum_{x,y} \hat{\delta}_t\cdot\frac{\partial\hat{\delta_t}}{\partial t} + \lambda_t \gamma_t \frac{\partial \hat{\mu}_t} {\partial t}    \nonumber \\
 &= \sum_{x,y} \hat{\delta}_t\cdot\bigg[\frac{\partial \mu_t}{\partial t} -\frac{\partial \hat{\mu}_t}{\partial t}\bigg]+\lambda_t \gamma_t \frac{\partial \hat{\mu}_t} {\partial t}           \nonumber \\
 &=\sum_{x,y} \hat{\delta}_t\frac{\partial \mu_t}{\partial t} -\hat{\delta}_t\frac{\partial \hat{\mu_t}}{\partial t} +\lambda_t \gamma_t \frac{\partial \hat{\mu}_t} {\partial t} \nonumber \\  
  &=\sum_{x,y} \underbrace{\hat{\delta}_t\frac{\partial \mu_t}{\partial t}}_{\leq 0} - \frac{\partial \hat{\mu_t}}{\partial t}\bigg[\hat{\delta}_t -\lambda_t \gamma_t \bigg] \nonumber \\  
  &\leq  - \frac{\partial \hat{\mu_t}}{\partial t}\bigg[\hat{\delta}_t -\lambda_t \gamma_t \bigg] \nonumber \\  
  &\leq 0 \nonumber
\label{eq:ProofII}
\end{align}
\begin{figure}[!t]
\begin{center}
\includegraphics[width=.98\textwidth]{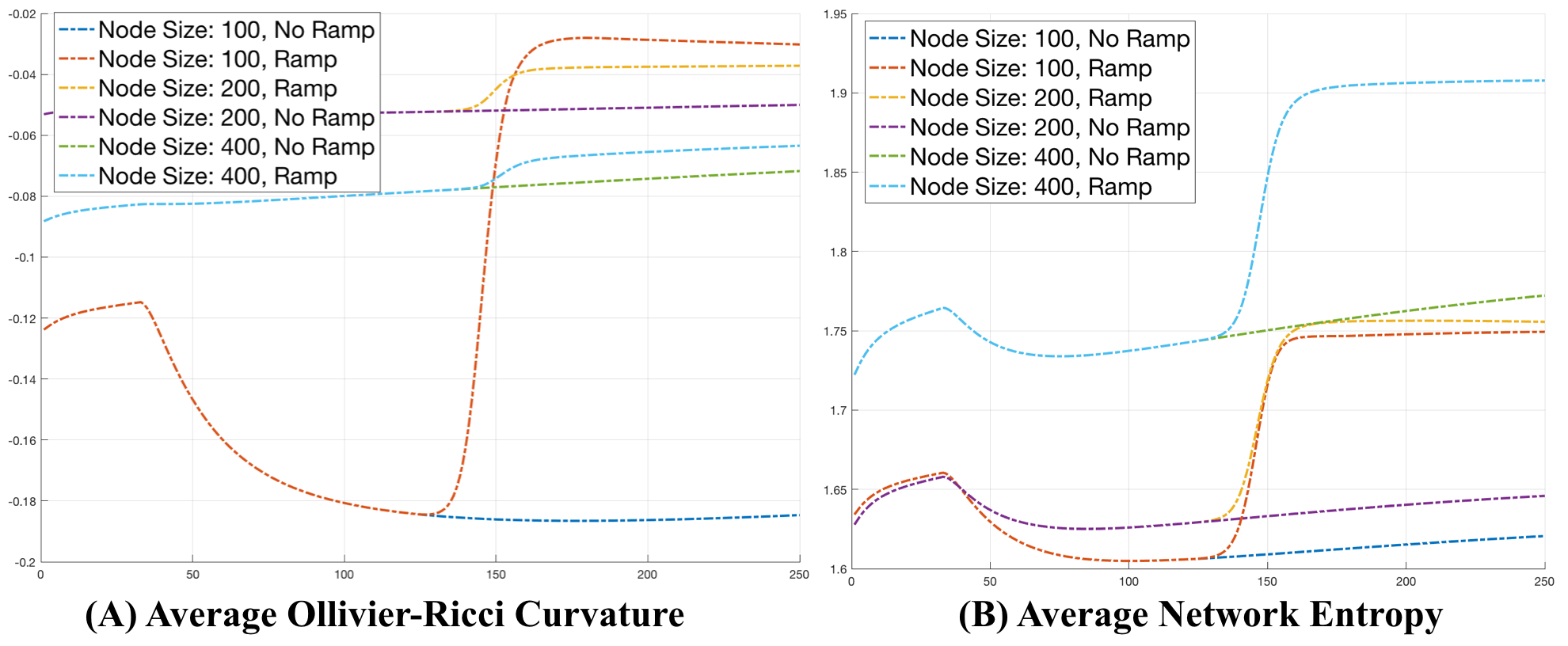}
\caption{We present results on scale-free networks of varying node sizes and the impact of operator input in ``injecting'' curvature of a single node associated with the highest topological degree.   (A) Average Ollivier-Curvature.  (B) Average Network Entropy.  Note:  Due to scaling, values for curvature and entropy differ; however, $\Delta H\times \Delta Ric\geq 0$}
\label{fig:rampup}
\end{center}
\end{figure}
As one can see from coupling both the estimator and autonomous model, a useful qualitative behavior emerges.  In particular, when the ``demon'' is satisfied with the agreement between $\mu_t(x,y)$ and their ideal $\mu^*(x,y)$ configuration, it is assumed that the total input $\lambda_t(x,y)$ will then remain constant.   That is, either the ``demon'' never needed to apply a correction or has otherwise stopped providing inputs.  Nevertheless, in this case, $\hat{\mu}^*_t(x,y)$ should ``follow'' $\mu_t(x,y)$.  On the other hand, when the total input error $\lambda_t(x,y)$ grows due to persistent input, $\hat{\mu}_t^*(x,y)$ will be increasingly driven towards $\lambda_t(x,y)$ irrespective of the agreement between $\hat{\mu_t}^*(x,y)$ and $\mu_t(x,y)$.  Ultimately, the demon has control of the seemingly accurate autonomous flow and can override systems actions.
\section{Results} 
In this section, we present results using graph curvature to indirectly control network entropy.   We caution the reader that these results are preliminary and to motivate theory presented.  This said, we conduct  experiments primarily focused involve scale-free networks as it provides natural topological hubs to test particular inputs can impede (induce fragility) via varying levels of input.  For all experiments, we generate networks via the Python NetworkX package and utilize the classic definition of network entropy \cite{Tesch1}.  

\begin{figure}[!t]
\begin{center}
\includegraphics[width=0.98\textwidth]{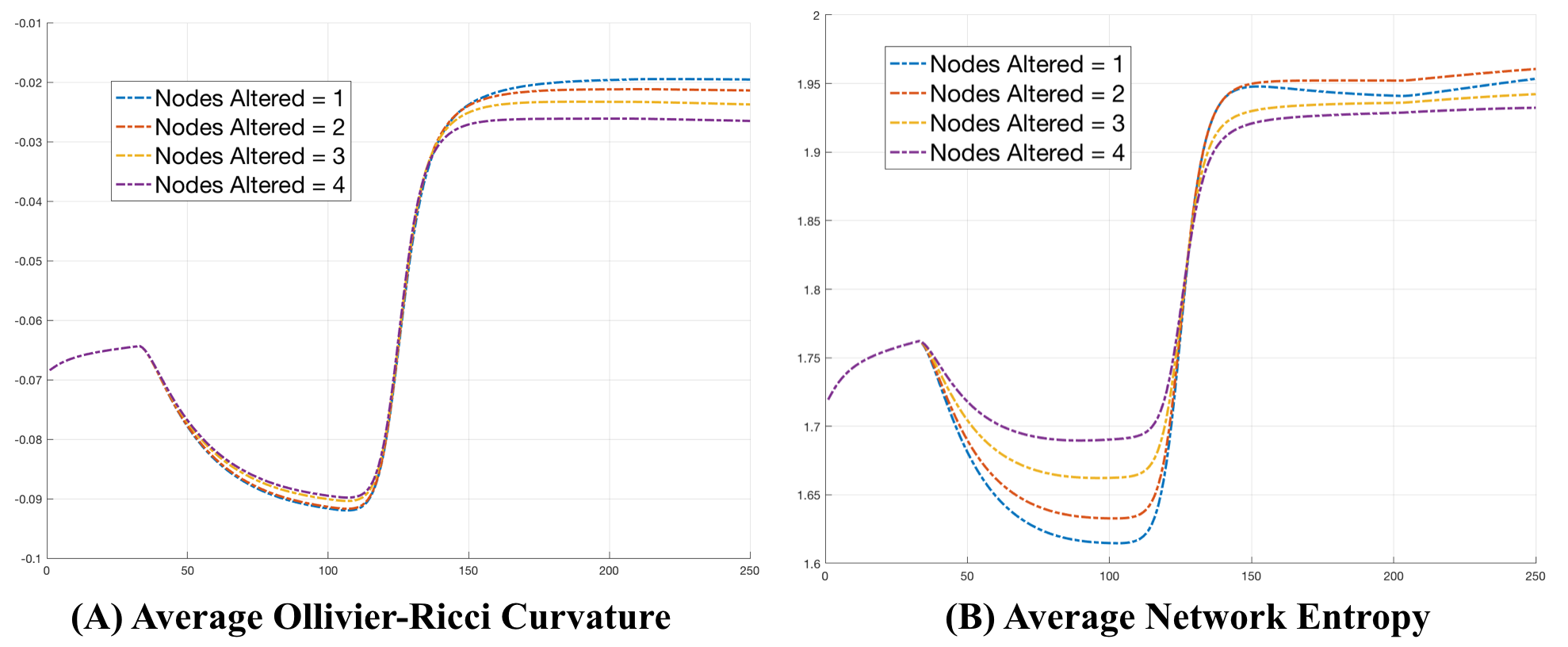}
\caption{We present results on scale-free networks of node size $n=400$ and the impact of providing operator input to several nodes associated with the highest topological degree.  (A) Average Ollivier-Ricci Curvature. (B) Average Network Entropy.  Note:  The number of nodes an operator interacts with naturally controls (increases/decreases) both curvature and entropy. }
\label{fig:nodes_alter}
\end{center}
\end{figure}

\begin{figure}[!t]
\begin{center}
\includegraphics[width=.90\textwidth]{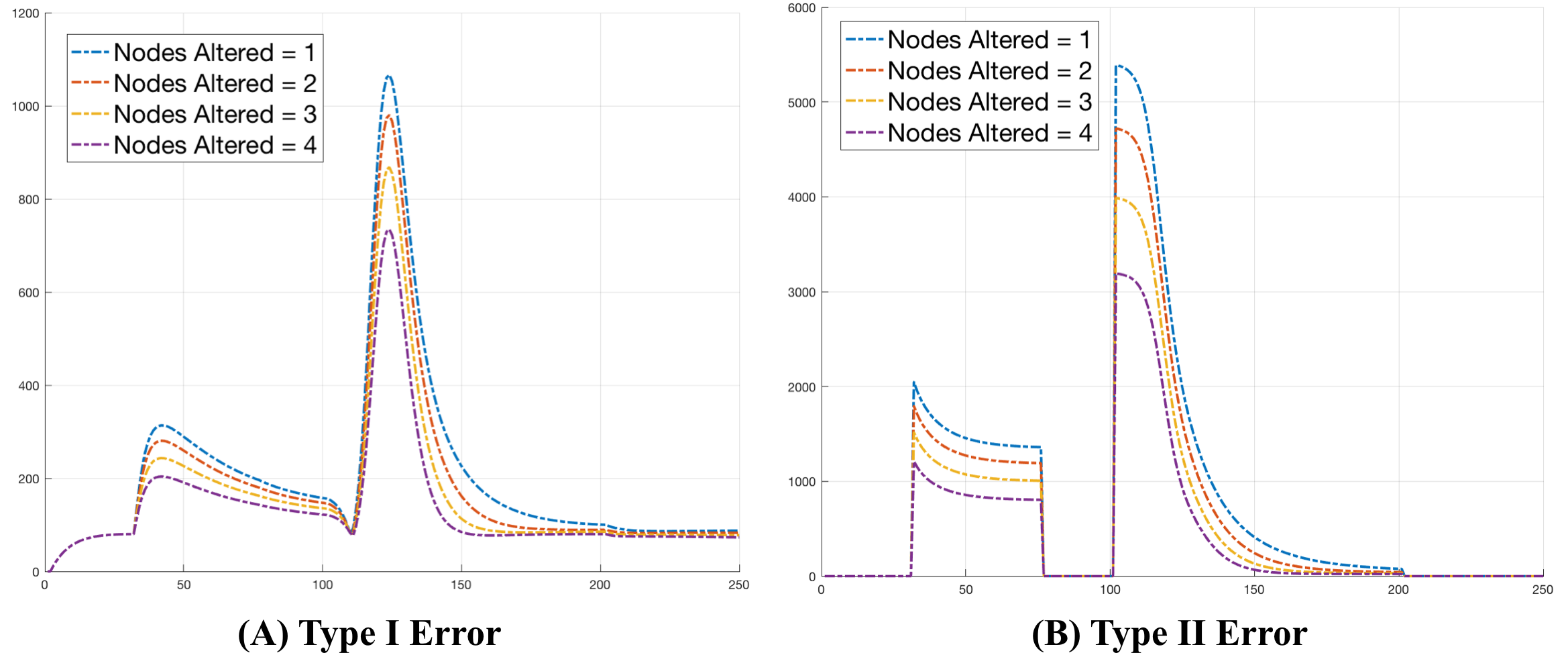}
\caption{We present errors results associated with Figure \ref{fig:nodes_alter} experiment related to node alteration and operator input.  (A) Type I Error (B) Type II Error.}
\label{fig:terrors}
\end{center}
\end{figure}

The first set of experiments focuses on controlling network entropy via discrete Ollivier-Ricci flow seen in Figure \ref{fig:rampup}.  As there exists an intimate connection that relates that changes in entropy are positively correlated with changes in Ricci curvature, i.e., $\Delta H \times \Delta Ric \geq 0 $, we generate scale-free networks with node sizes of $n=[100, 200,400,600]$ with uniform edge weights.  From this, we target the node with the highest degree and begin ``injecting'' input and allow for our flow to evolve as described by the coupled feedback equations in equation (9) and equation (15).  To be more precise, at time $t=[30, 75, 120, 175]$ we make an input of values $p=[-2, 2, 4, -4]$, respectively.  The resulting changes in network entropy as well as average Olliver-Ricci curvature can be seen as solid colored lines in Figure \ref{fig:rampup}.  Remarkably, we see a very close relationship between network entropy and that of network curvature.  Furthermore, to validate our ability to ``change direction'' in terms of altering network entropy, we re-run the same experiment with a slight change by ``turning off'' input at $t=120$; i.e., for $t=[30, 75, 120, 175]$ we make an input of values $p=[-2, 2, 0, 0]$, respectively.  Once again, we see the natural impact and differences of operator input.
\begin{figure*}[!t]
\begin{center}
\includegraphics[width=0.98\textwidth]{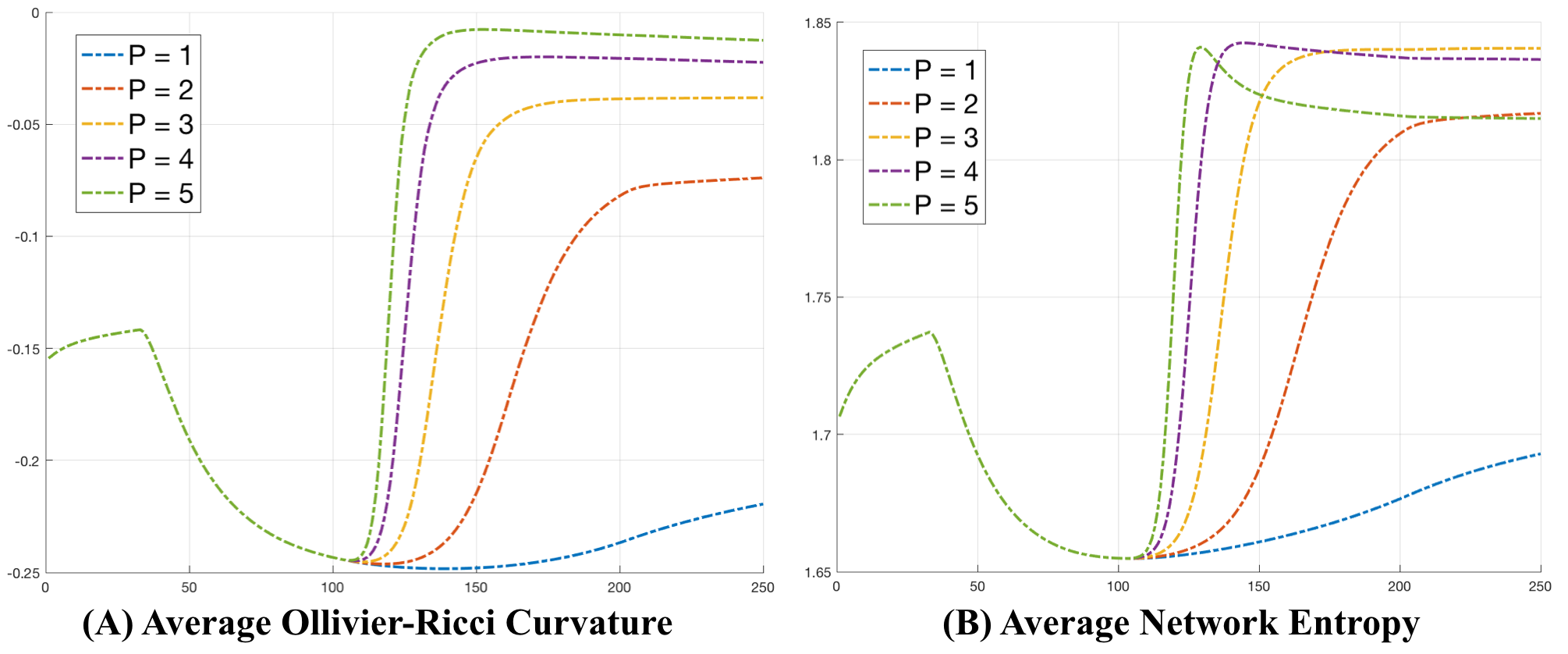}
\caption{We present results on scale-free networks of node size $n=200$ and the impact of operator input at varying levels of $p$ from $t=100$ to $t=200$ for the node associated with the highest topological degree.  (A) Average Ollivier-Ricci Curvature. (B) Average Network Entropy.  Note:  The degree of operator input naturally controls (increases) both curvature and entropy in the aforementioned time region. }
\label{fig:pvalues}
\end{center}
\end{figure*}
On the other hand, we also want to measure how the degree of input (e.g., choosing the constant $p$) alters networks entropy as well as the impact of altering more than one hub node in a given network.  To this end, we generate scale-free networks of node size $n=200$.  From this, at time $t=[30, 75, 100, 200]$ we make an input of values $p=[-2, 2, \theta, -\theta]$ where $\theta = [5,4,3,2,1]$.   As one can see from Figure \ref{fig:pvalues}, we see exactly this behavior which also correlates to the degree of operator input.  Next, we make a slight alteration to this experiment and now at iterations [30, 75, 100, 200], we make an operator input of values $p=[-2, 2, 4, -4]$ similar to the first experiment for a scale-free network of node size $n=400$.  However, we now plot changes in network entropy and network curvature as a function of altering the top $n$ nodes with the highest degree.  Again, we see the behavior that is to be expected in increasing network robustness as seen in Figure \ref{fig:nodes_alter}.  For this experiment, Figure \ref{fig:terrors} shows Type I and Type II error for completeness.
\section{Conclusions and Future Work}
We propose a network control framework that couples the discrete Ollivier-Ricci flow with operator input from a feedback perspective.  To this end, we provide the necessary stability conditions in the sense of Lyapunov.  This said, there exists several avenues that we are currently pursuing.  The above framework has potential biological application towards the real-world setting in which we often seek to understand how can induce fragility on targets that are deemed ``undruggable'' \cite{Sandhu1}.  We also aim to extend the above framework for non-constant user input, time-delayed response, and as applied to specific application domains. As such, this work has laid the foundation for which further examination is needed.

\end{document}